\documentstyle[epsfig,prl,twocolumn,aps,floats]{revtex}
\begin{document}
\draft
\wideabs
{\title{Scanning Tunneling Microscopy of a Luttinger Liquid}
\author{Sebastian Eggert}
\address{Institute of Theoretical Physics,
Chalmers University of Technology and G\"oteborg University,
S-412 96 G\"oteborg, Sweden}
\date{Submitted: August 2, 1999.  Last change: \today}
\maketitle
\begin{abstract}
Explicit predictions for Scanning Tunneling Microscopy (STM) on 
interacting one-dimensional electron systems are made using 
the Luttinger liquid formalism.  The STM current changes with distance 
from an impurity or boundary in a characteristic way, which reveals the 
spin-charge separation and the interaction strength in the system.
The current exhibits Friedel-like oscillations, 
but also carries additional modulated behavior as a function of 
voltage and distance, which shows the spin-charge separation in 
real space.  Moreover, very close to the boundary
the current is strongly reduced, which is an indication of the
interaction strength in the system.
\end{abstract}
}
\narrowtext

In the last two decades the interest in quasi one-dimensional physics has 
been spurred by experimental progress in constructing smaller and more
refined structures such as carbon nanotubes, atomic point 
contacts, and mesoscopic quantum wires produced by etching\cite{etched} 
or cleaved edge overgrowth\cite{cleaved}.
More recently it was even possible to produce single atomic chains by 
depositing gold on a silicon surface\cite{baer}.

The theoretical foundation for describing 
interacting one-dimensional electrons was laid in the early 80's
with the concept of a Luttinger Liquid (LL)\cite{haldane}. Interestingly, the
electron-electron interactions cannot ever be neglected in one dimension
which makes those systems fundamentally non-Fermi-liquid like.  
The elementary excitations are described by separate spin and
charge quasi-particles which move at different velocities\cite{VOIT}.  
The correlation functions are power-laws with exponents 
that are related to a single interaction constant $g$.  

Angle resolved photo-emission experiments have made some progress in 
identifying a possible signature of spin-charge separation in quasi 
one-dimensional compounds\cite{baer,photo}. 
On the other hand, in mesoscopic wires 
most experiments focus on conductivity measurements\cite{etched,cleaved}.
However, from those experiments it is very difficult 
to extract information about the fundamental 
interactions within the wire.  
To test the
important theoretical concept of spin-charge separation 
in mesoscopic systems other methods must be considered.  

One difficulty in producing a good wire is the fact that even small 
impurity perturbations effectively cut the wire 
at low temperatures\cite{kane}. However, such boundaries 
give rise to other interesting effects, 
which can even reveal the LL behavior as we will show below. 
One well-known impurity effect in metals is
the induced charge density fluctuation at twice 
Fermi wave-vector, the so-called Friedel oscillation. 
In the case
of carbon nanotubes the Friedel oscillations have already been used to
show the more complicated electronic structure\cite{mele}, 
which stems from a rolled up two dimensional graphite sheet.  

We now make predictions for an STM experiment along a 
simple quantum wire described by the LL model
with an open end instead of leads  
i.e.~electrons which have been
confined to move in one dimension by clever gating or an appropriate 
deposit on a surface. 
We show that 
the spatial structure of the tunneling current reveals both the 
spin-charge separation and the interaction strength in the LL system.

The STM current $I$ is directly related to how many electron states
are locally available in the LL system and in the tunneling tip.  In particular,
at position $r$ and for a given tunneling voltage $V$ we can write
\begin{equation}
I(V, r) \propto \int_0^V d\omega  \ N(\omega, r) f(\omega-V), \label{current}
\end{equation}
where $N(\omega, r)$ is
the local density of states (DOS) in the LL with $\omega$
measured relative to the Fermi energy, and $f$ is the DOS
in the tip.  We do not know the detailed properties of the tip,
but we can assume that $f(\omega)$ is smooth compared to the more singular
structure of $N(\omega)$, so that $f=\rm const$ is a valid 
approximation.  The DOS of the system is given in terms of the 
time-time correlation functions in the LL at position $r$ 
\begin{equation}
N(\omega,r)  \equiv  \frac{1}{2 \pi} \int_{-\infty}^\infty
\!\!e^{i \omega t} \langle\left\{
\Psi^{}_\sigma(r,t),
\Psi^{\dagger}_\sigma(r,0)
\right\}\rangle dt,
\label{N_omega}
\end{equation}
where $\omega$ is measured relative to the Fermi energy.
It is already well understood how to calculate 
the DOS as a function of energy $\omega$ with the 
LL formalism\cite{VOIT,meden}, 
but as described in this Letter it is the spatial 
structure as a function of distance $r$ from a boundary that also 
explicitly shows the spin-charge separation.

The LL Hamiltonian describes one-dimensional electrons 
with short range interactions 
in the low-temperature limit below some energy cut-off $\Lambda$. 
In that limit, the dispersion is approximately linear and the 
electron-electron scattering rate can be taken as momentum independent.
The high energy cut-off $\Lambda$ is large compared to the temperature,
but about one magnitude less than the 
bandwidth for typical lattice models. 
The electron field $\Psi_{\sigma}(x)$ is expressed
in terms of left- and right-moving Fermions at the Fermi points $\pm k_F$ 
\begin{equation}  
\Psi_{\sigma}(x) = 
e^{-ik_Fx}\psi_{L}^{\sigma}(x) + e^{ik_Fx}\psi_{R}^{\sigma}(x). \label{lin}
\end{equation}
We also define the Fermion currents 
$J_{L/R}^\sigma \equiv :\!\psi_{L/R}^{\sigma \ \dagger}\psi_{L/R}^{\sigma}\!:$.
Umklapp scattering is suppressed away from half filling, and remarkably all 
forward scattering processes can be described by expressing the
spin and charge currents in terms of separate bosonic variables
$\phi_c$ and $\phi_s$ and their conjugate momenta $\Pi_c$ and $\Pi_s$
\begin{equation}
J_L^{c/s} \equiv \case{1}{\sqrt{2}} \left( J_L^\uparrow \pm
J_L^\downarrow \right) =  \case{1}{\sqrt{4 \pi}} \left(\Pi_{c/s} +
{\partial_x \phi_{c/s}}\right)
\end{equation}
The LL Hamiltonian density is then written as
\begin{equation}
{\cal H}  \ =  \     \case{v_c}{2}
\left[g^{-1}({\partial_x \phi_c})^2 +  g \Pi_c^2\right] \ + \
\case{v_s}{2} \left[({\partial_x \phi_s})^2 +  \Pi_s^2\right]
\end{equation}
which describes two independent bosonic excitations for spin and charge
separately.  Due to SU(2) invariance the spin boson is a free field,
but the charge boson gets rescaled by the LL parameter 
$g$ which is less than unity for repulsive interactions.

After the linearization around the Fermi points in Eq.~(\ref{lin})
we can write (omitting the spin indices $\sigma$)
\begin{eqnarray}
&&\langle \Psi^{}(r,t) \Psi^{\dagger}(r,0)\rangle   =  
\langle \psi_{L}^{}(r,t)\psi_{L}^{\dagger}(r,0) \rangle +
\langle \psi_{R}^{}(r,t)\psi_{R}^{\dagger}(r,0) \rangle  \nonumber \\
&&   + \ 
e^{i 2k_F r} \langle \psi_{R}^{}(r,t)\psi_{L}^{\dagger}(r,0) \rangle +
e^{-i 2k_F r} \langle \psi_{L}^{}(r,t)\psi_{R}^{\dagger}(r,0) \rangle .
\label{GF}
\end{eqnarray}
The last two terms carry a rapid oscillation with twice
the Fermi wave-vector $2 k_F r$ reminiscent of Friedel-oscillations, 
while the first two terms are slowly varying.  The tunneling current in
Eq.~(\ref{current}) therefore has a rapidly oscillating
Friedel part $I_{\rm osc}$ and a uniform part $I_{\rm uni}$
with a spatial dependence that is smooth compared to $e^{i 2 k_F r}$
\begin{equation}
I(V,r) = I_{\rm uni}(V,r) + \cos(2 k_F r + \phi) I_{\rm osc}(V,r).
\end{equation}
In a translational invariant system 
the left- and right-movers are uncorrelated 
$\langle \psi_{L}^{}\psi_{R}^{\dagger} \rangle = 0$
and we cannot observe any spatial structure.
A generic impurity, however, scatters left- into right-movers and 
the resulting correlation functions depend  
on the distance $r$ from the end.  Such boundary correlation
functions have first been calculated for the spin-channel\cite{eggert} and 
later also for the full electron field\cite{fabrizio,boundcorr,mattsson}.  
We consider an open boundary at the origin $r=0$ of a relatively
long system so that $N(\omega)$ is continuous 
(the other end of the system is far away and can be 
neglected for now).  In that case we find for the uniform terms
\begin{eqnarray}
\langle \psi_{L}^{}(r,t)\psi_{L}^{\dagger}(r,0) \rangle 
& \propto & \left[\frac{1}{\alpha + i v_s t}\right]^{\slantfrac{1}{2}} 
\left[\frac{1}{\alpha + i v_c t}\right]^{a+b}
\nonumber \\ &\times &
 \left[\frac{4r^2}{(\alpha + i v_c t)^2+4 r^2}\right]^{c} \label{LL}
\end{eqnarray}
and for the Friedel terms
\begin{eqnarray}
\langle \psi_{R}^{}(r,t)\psi_{L}^{\dagger}(r,0) \rangle 
&\propto & \left[\frac{1}{\alpha + i (v_s t - 2 r)}\right]^{\slantfrac{1}{2}}
 \left[\frac{1}{\alpha + i (v_c t - 2 r)}\right]^a
\nonumber \\ &\times &
 \left[\frac{1}{\alpha + i (v_c t + 2 r)}\right]^b
 \left[\frac{|2r|}{\alpha + i v_c t}\right]^{2 c} \label{RL}
\end{eqnarray}
where the exponents are given in terms of the interaction parameter $g$
$$
a  =  (\case{1}{g} + g + 2)/8 
, \ b = (\case{1}{g}+g-2)/8,  \ c = (\case{1}{g}-g)/8.
$$
The short distance cut-off $\alpha \sim v/\Lambda$ is 
small compared to all other length-scales in the system, and we take
$\alpha \to 0^+$ in all following calculations.  The corresponding expressions 
for left- and right-movers exchanged can be obtained by taking $r \to -r$.

\begin{figure}
\begin{center}
\mbox{\epsfig{file=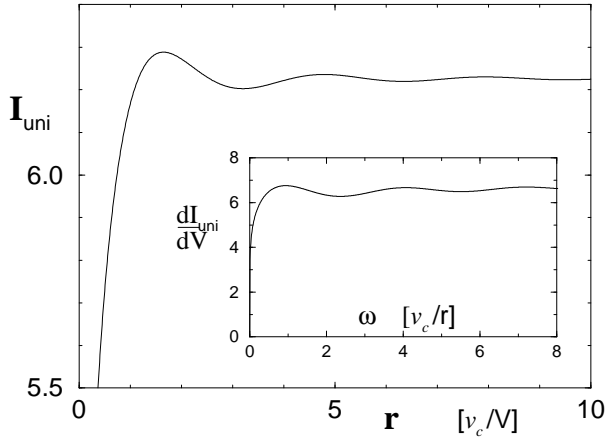,width=3.05in,angle=0}}
\end{center}
\caption{The uniform current $I_{\rm uni}$ in arbitrary units
as a function of $r$ for $g=3/4$.  
The inset shows $dI_{\rm uni}/dV$ 
as a function of $\omega$ at a position $r$.}
\label{uni}
\end{figure}

Let us first consider the uniform part of the current $I_{\rm uni}$
as determined by the analytic structure of Eq.~(\ref{LL}). 
A change of variables $t' = t/r$  in Eq.~(\ref{N_omega}) and 
$\omega' =  r \omega$ in Eq.~(\ref{current}) shows that the uniform 
current $I_{\rm uni}$ is a function of the scaling variable $rV$
\begin{equation}
I_{\rm uni}(V, r) = r^{-(1/g+g+2)/4} F(r V) \label{scaling}
\end{equation}
For non-interacting electrons ($g=1, \ v_c = v_s = v$) we get 
$a=\case{1}{2}, \ b=c=0$ corresponding to a single pole
of order one at $t=i\alpha$ in Eq.~(\ref{LL}).
The integration in Eq.~(\ref{N_omega}) gives a constant from the residue
in the upper half plane, i.e. the DOS is
independent of $r$ and $\omega$, and Eq.~(\ref{current}) simply
gives $I_{\rm uni} \propto V$ for non-interacting electrons.  However, even for
small interactions the single pole splits into three singularities  
at $t= i\alpha$ and $t = \pm 2 r/v_c + i \alpha$ in Eq.~(\ref{LL}). 
Close to the boundary the behavior of the Fourier
transform in Eq.~(\ref{N_omega}) is then dictated by the large time behavior 
of Eq.~(\ref{LL}) and we find 
\begin{equation}
I_{\rm uni} \propto r^{(1/g-g)/4} V^{(1/g+1)/2} \ \ {\rm for} \ \  r < v_c/V 
\label{deplete}
\end{equation}
i.e. a characteristic depletion as $r\to 0$ for repulsive interactions 
$g < 1$.  On the other hand far away from the boundary $r \gg v_c/V$
the behavior is dominated by the most divergent 
singularity and we find 
\begin{equation}
I_{\rm uni} \propto V^{(1/g+g+2)/4} \ \ {\rm for} \ \  r > v_c/V,
\end{equation}
which is largely independent of $r$.  
However, the integration of the deformed contour 
in Eq.~(\ref{N_omega}) around the branch-cuts of the weaker singularities
also contributes, multiplied by a corresponding 
slowly oscillating ``residue-factor'' $e^{\pm 2 i r \omega/v_c}$. 
This results in an additional slowly oscillating contribution with
$\cos( 2  r V/v_c)$ that drops off with
$r^{(1/g -g-8)/8}$.  
The depletion with the slow oscillations towards a constant current 
is depicted in Fig.~(\ref{uni})
for $g=3/4$ from doing the integrals numerically.
None of the singularities in Eq.~(\ref{LL}) 
depend on the spin velocity $v_s$ and  therefore
the uniform current $I_{\rm uni}$ will {\it not} show
any signs of spin-charge separation. Nonetheless the weaker singularities 
still give the characteristic slowly oscillating 
structure due to interactions.

\begin{figure}
\begin{center}
\mbox{\epsfig{file=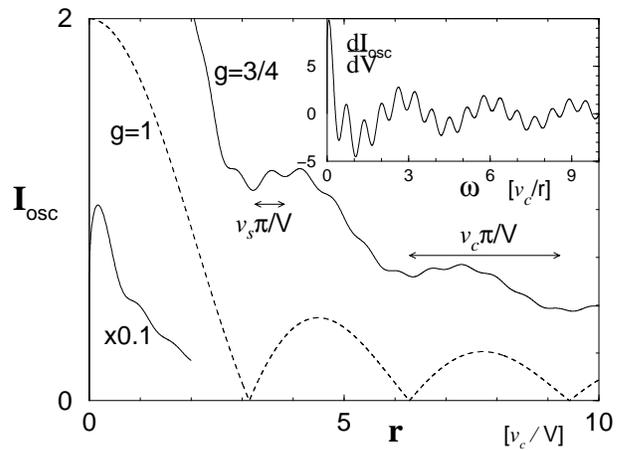,width=3.05in,angle=0}}
\end{center}
\caption{The Friedel amplitude of the tunneling current in arbitrary units
as a function of $r$ for $g=1$ and $g=3/4,\ v_c/v_s = 5$.  
The inset shows $dI_{\rm osc}/dV$ at a 
position $r$ with an arbitrarily chosen phase $e^{i 2 k_F r}=1$.}
\label{osc}
\end{figure}

The effect of the last two ``Friedel'' terms in 
Eq.~(\ref{GF}) on the other hand will reveal the spin-charge separation.
The amplitude $I_{\rm osc}$ of the rapidly oscillating Friedel current 
has the same scaling form as in Eq.~(\ref{scaling}) but 
with an even richer behavior for $F$.  Already for non-interacting electrons
we find a single pole at $t = 2 r/v + i \alpha$ in Eq.~(\ref{RL})
which results in $N(\omega,r)
\propto \cos(2 k_F r + 2 r \omega/v)$.  The integration in 
Eq.~(\ref{current}) gives a strong $r$-dependence of the
amplitude $I_{\rm osc} \propto \sin(\frac{r V}{v})/r$ 
(``amplitude modulation'') as shown in Fig.~(\ref{osc}) for $g=1$.  
With interactions we now find four different singularities at 
$t=i \alpha, \ t = \pm 2 r/v_c + i \alpha$ and $t = 2 r/v_s + i\alpha$ 
in Eq.~(\ref{RL}). The long-time behavior of  Eq.~(\ref{RL}) is the same 
as for the uniform terms in Eq.~(\ref{LL}),
so that we get the same universal depletion for 
$I_{\rm osc}$ as in Eq.~(\ref{deplete}) 
very close to the boundary $r< v_s/V$.
For larger distances from the boundary $r \gg v_c/V$, however, the 
drop-off of the Friedel current is determined by the leading singularity
in Eq.~(\ref{RL})
\begin{equation} 
I_{\rm osc} \propto  \left\{ \begin{array}{lcl}
               r^{-(1/g+g+2)/8} V^{(1/g+g+2)/8} & & 
               \frac{1}{3} < g <1\\
               r^{-(1+g)/2} V^{(1/g-g)/4} && g < \frac{1}{3}
                  \end{array}
           \right. .
\end{equation}
More importantly, the Friedel amplitude $I_{\rm osc}$ has an oscillating  
superstructure from the residue factor of each singularity. The strong 
amplitude modulation with $\sin\frac{rV}{v}$ has already been demonstrated 
for the non-interacting case, but the ratio of the velocities
$v_c/v_s$ can now be much larger than one.  Therefore, we observe 
two separate spin and charge amplitude modulations 
of $I_{\rm osc}$ with $r V/v_s$ and with $r V/v_c$ respectively 
(which are still 
smooth compared to the overall oscillation of $2 k_F r$).  This behavior is
demonstrated in Fig.~(\ref{osc}) for $g=3/4$ and $v_c/v_s = 5$.
The physical interpretation is that we observe the superposition 
of all electron wave-functions 
in the energy range from $0$ to $V$ which exhibits the 
spin-charge separation due to the interference from the boundary.  

We have shown that the spatial electronic structure of an LL 
indeed shows the signatures of the spin-charge separation, but it 
is important to critically analyze to what extent this could be
observed in a realistic STM experiment.  
As an example we consider the monoatomic  
gold chains on a slanted silicon surface Si(111)-Au(5x1),
which showed the signature of LL behavior in photo-emission
experiments\cite{baer}.  The spin-charge separation seems to be present
for all excitations over the entire bandwidth (ca.~1 eV), but to observe
the particular STM structures that we predict here, the voltage has 
to be below the cut-off $\Lambda$ which is about 0.1 eV for this system.
The most easily observable feature in an STM experiment is probably
the depletion of the tunneling current as a function of distance near 
the boundary.  The range of this depletion is given by $r \approx v_F/V$, where 
$v_F \approx 6\times 10^5 \rm \case{m}{s}$ for the gold chains, so that 
for $V < 100 \rm meV$ the range
is at least $r \agt 40  \rm \AA$.  Already from the shape of this depletion an
approximate estimate of $g$ can be made from Eq.~(\ref{deplete}).
Secondly, it is important to analyze the 
Friedel oscillations, which is a more difficult task.   For the gold
chains the Fermi vector is $k_F = \pi/2a$ with an interatomic spacing
of $a= 3.83 \rm \AA$, so that the $2 k_F r$ oscillations are commensurate
with the lattice.  This makes the Friedel oscillations easier to 
detect, but the small amplitude modulations in Fig.~(\ref{osc}) 
may not show up very clearly.  The Friedel amplitude is modulated
by at most 30\% by the charge waves, while the spin modulations 
are even smaller.  The spin modulations are actually {\it weaker} 
for stronger interactions 
(about 10\% for $g\sim 1$, but only 1\% for $g < 1/2$).  Nonetheless, even
if only a small hint of those superstructures shows up in an STM image,
it should be possible to track those modulations for different 
voltages. This will change their period systematically and may 
make it possible to identify them unambiguously. 
Interestingly, the total modulation is much {\it stronger} 
without any spin-charge separation (g=1), so if it cannot 
be observed at all it would also be an indication of interaction effects.
For larger voltages  $V > \Lambda$ it is more instructive to look 
at the total density of electrons given 
directly by Eqs.~(\ref{LL}-\ref{RL}) in the limit $t\to 0$, 
which recovers the established drop-off of the Friedel 
terms with $r^{-(1+g)/2}$\cite{fabrizio,LLfriedel}, but 
without any spin-charge modulations.  

The measurement of $dI/dV$ as a function of voltage 
in an STM experiment (spectral
mode) will give additional information about the local DOS.  Close to the
boundary $dI/dV$ has considerable structure as indicated in the insets
of Figs.~(\ref{uni}-\ref{osc}). The additional 
contribution $dI_{\rm osc}/dV$ from the Friedel terms 
in Fig.~(\ref{osc}) is smaller and depends 
on the choice of phase $e^{i 2 k_F r}$, but
shows the separate spin-charge effects very clearly.
The gold chains mentioned above showed 
a surprisingly strong depletion of the spectral weight at the 
Fermi surface as a function of energy 
in photo-emission experiments\cite{baer}.  
It would be interesting to see if STM experiments on the very
same samples also show such a strong depletion of the DOS
with a characteristic power-law $V^{2 b}$. 

One approximation we made is that the DOS
in the tip $f(\omega)$ is relatively smooth, which is 
not too restrictive since the observed structures 
in Figs.~(\ref{uni}-\ref{osc}) can be reproduced
for almost all forms of $f(\omega)$ as long as there is a
sharp upper limit at $V$ in the integral of Eq.~(\ref{current}). 
It is also possible to average over both signs of the voltage
to eliminate the DOS of the tip somewhat,  since the anticommutator 
in Eq.~(\ref{N_omega}) gives the same singularities for $t \to -t$, 
i.e.~$N(\omega) = N(-\omega)$.  

The distance from the tip to the sample is also important,
since the tip may influence the system and play the role of an impurity 
itself.  The optimal distance can be determined in the actual experiment
by operating in spectral mode somewhere in the bulk of the wire.
If the $dI/dV$ curve changes qualitatively as the tip-sample distance 
is decreased (except for an overall scale), then this would be a clear sign 
that the tip influences the sample.  In particular, if the tip 
starts to act as an impurity 
the $dI/dV$ curve should show a stronger depletion with the  
power-law $V^{(1/g-1)/2}$ as $V\to 0$.

So far we have only considered a perfectly reflecting impurity, because
it is expected that any generic perturbation
renormalizes to the open boundary fixed point\cite{kane}. 
However, in an intermediate range around a ``weaker''  
impurity the Friedel effects show a non-universal behavior described in
terms of a form factor\cite{LLfriedel}.  
The range of this so-called boundary 
shrinks to zero for a perfectly reflecting barrier. Also  
interesting are ``active'' impurities that have a net magnetic moment 
or carry an electric charge.  In the presence of interactions those impurities
may be overscreened, i.e.~the nearest electrons overcompensate for the
impurity charge or spin and in turn get screened by the next-nearest 
neighbors, etc.  This finally results in a screening cloud which is also a 
$2 k_F r$ effect, since the impurity Hamiltonian $H_{\rm imp}$  
induces a non-zero expectation value in the Friedel terms of Eq.~(\ref{GF}), 
i.e.~$\langle \psi_{L}^{}\psi_{R}^{\dagger} H_{\rm imp} \rangle \neq 0$.
The presence of both backscattered and induced Friedel terms has recently 
been demonstrated for a two-channel Kondo impurity\cite{2CK}.

Finally, we must also consider the effect of a second boundary in a
finite system at $r=L$. The correlation functions 
in Eqs.~(\ref{LL}-\ref{RL}) are then described by powers of  
sine functions $\sin\frac{\pi v t}{2 L}$\cite{fabrizio,mattsson}.  
We expect that
the spatial structure from the interference of the standing waves gives 
a similar picture as in Figs.~(\ref{uni}-\ref{osc}) close to the 
boundaries as long as $V \gg \pi v/L$. However, a more dramatic 
finite size effect is a discrete spectrum $N(\omega)$ due to the 
appearance of $\delta$-functions in Eq.~(\ref{N_omega}).  This results in
Coulomb-blockade-like charging steps in $I(V)$,
which can also reveal the spin-charge separation and the interaction
strength in the LL\cite{mattsson}.

In conclusion we have shown that the tunneling
current has decaying Friedel-like oscillations in a range around
a boundary, but additionally the Friedel amplitude carries a characteristic
periodic modulation in real space 
which reveals the separate spin and charge parts of
the electron wave-functions.  The period of those modulations is 
a function of the tunneling voltage, which is assumed to be small.
We also find a characteristic depletion very close to the boundary of both
the Friedel current and the uniform current, which is
immediately related to the interaction constant in the wire.

The author is very grateful to Henrik Johannesson for 
inspirational discussions and to Yves Baer for an early copy
of Ref.~\onlinecite{baer}.  This research was supported
by the Swedish Natural Science Research Council with the 
grants F-AA/FU 12288-301 and S-AA/FO 12288-302.  



\begin{thebibliography}{19}

\bibitem{etched} S.~Tarucha {\it et al.}, 
Solid State Comm. {\bf
94}, 413 (1995).
\bibitem{cleaved} A.~Yacoby {\it et al.},
Phys. Rev. Lett. {\bf 77}, 4612 (1996).
\bibitem{baer} P.~Segovia {\it et al.}, 
Nature {\bf 402}, 504 (1999). 
\bibitem{haldane} F.D.M.~Haldane, { J. Phys. C: SSP}
{\bf 14}, 2585 (1981).
\bibitem{VOIT} For a review see J. Voit, 
{ Rep. Prog. Phys.} {\bf 58}, 977 (1995).
\bibitem{photo} C.~Kim {\it et al.},
Phys. Rev. Lett. {\bf 77}, 4054 (1996).
\bibitem{kane} C.L.~Kane, M.P.A.~Fisher, Phys. Rev. B {\bf
46}, 7268 (1992).
\bibitem{mele} C.L.~Kane, E.J.~Mele, Phys. Rev. B {\bf 59}, 12759 (1999); 
W. Clauss 
{\it et al.}, Europhys. Lett. {\bf 47}, 601 (1999).
\bibitem{meden} K.~Sch\"onhammer,  V.~Meden, { Phys. Rev. B} 
{\bf 47}, 16205 (1993); J. Voit,
{ J. Phys. C: CM} {\bf 5}, 8305 (1993).
\bibitem{eggert} S. Eggert, I. Affleck, {Phys. Rev. B} {\bf 46}, 10866 
(1992); { Phys. Rev. Lett.} {\bf 75}, 934 (1995). 
\bibitem{fabrizio}  M.~Fabrizio, A.O.~Gogolin, 
 { Phys. Rev. B} {\bf 51}, 17827 (1995).
\bibitem{boundcorr} S.~Eggert {\it et al.}, 
Phys.~Rev.~Lett. {\bf 76}, 1505 (1996).
\bibitem{mattsson} A.E.~Mattsson {\it et al.}, 
Phys.~Rev.~B {\bf 56}, 15615 (1997);  S.~Eggert {\it et al.},
Phys.~Rev.~B {\bf 56}, R15537 (1997).
\bibitem{LLfriedel} R.~Egger, H.~Grabert, Phys. Rev. Lett. {\bf 75}, 3505 
(1995); A.~Leclair {\it et al.}, 
Phys. Rev. B {\bf 54}, 13597 (1996).
\bibitem{2CK} S.~Eggert, S.~Rommer, Phys. Rev. Lett. {\bf 81}, 1690 (1998); 
Physica B {\bf 261}, 200 (1999).
\end{thebibliography}
\end{document}